# The WorldGrid transatlantic testbed: a successful example of Grid interoperability across EU and U.S. domains

F. Donno
*CERN, CH-1211 Geneva 23, Switzerland*

V. Ciaschini
*INFN-CNAF, Viale Berti Pichat 6/2, 40127 Bologna, Italy*

D. Rebatto, L. Vaccarossa
*INFN,Via Celoria 16, 20133 Milano, Italy*

M. Verlato
*INFN,Via Marzolo 8, 35131 Padova, Italy*

The European DataTAG project has taken a major step towards making the concept of a worldwide computing Grid a reality.  In collaboration with the companion U.S. project iVDGL, DataTAG has realized an intercontinental testbed spanning Europe and the U.S. integrating architecturally different Grid implementations based on the Globus toolkit.
WorldGrid is an intercontinental testbed running a given set of core services and several collective optional services, sharing the same authentication and authorization mechanism, and using a common schema for resource location and status in the context of a Virtual Organization. The WorldGrid testbed has been successfully demonstrated at SuperComputing 2002 and IST2002 where real HEP application jobs were transparently submitted from U.S. and Europe using "native" mechanisms and run where resources were available, independently of their location.
In this paper we describe the architecture of the WorldGrid testbed, the problems encountered and the solutions taken in realizing such a testbed. In particular, we describe the problems related with the authentication and authorization mechanism; the job submission mechanism that allows for user requirement match-making through a EU DataGrid GLUE-aware Resource Broker successfully extended to U.S. sites; replica management and registration for CMS applications through DataGrid middleware; and finally creating a coherent and automatically installable software distribution that can satisfy existing deployment tools such as LCFG used by DataGrid in Europe and PACMAN used by VDT in U.S.. In order to monitor the behavior and performance of WorldGrid, the EDT-Monitor package has proven to be a good prototype to be deployed at operation centers. Through a geographical map of the sites, an operator has access to the resource and services status and can spot problems on-line.
With our work we present an important step towards interoperability of Grid middleware developed and deployed in Europe and the U.S.. Some of the solutions developed in WorldGrid will be adopted by the LHC Computing Grid first service. To the best of our knowledge, this is the first large-scale testbed that combines middleware components and makes them work together.

**THCT006**

## 1. INTRODUCTION

The IST2002 [6] conference in Copenhagen and SC2002 [7] in Baltimore (U.S.) hosted successful demonstrations of the WorldGrid transatlantic testbed where HEP application jobs were transparently submitted through Grid Web portals for execution on resources spread across Europe and the U.S.. This was the result of an initiative started in collaboration between the European DataTAG (EDT) [1] and the iVDGL [2] projects. The goal was to prove the feasibility of a worldwide Grid based on existing, architecturally different Grid infrastructures running the same basic core services provided by the Globus Toolkit [5] and offering transparent access to the distributed computing infrastructure necessary to "data-intensive" modern applications.

The motivations driving this effort were:
- Start operative collaborations between European and U.S. Grid projects.
- Show interoperability between Grid domains for applications submitted by users from different virtual organizations.
- Demonstrate controlled use of shared resources subject to agreed policy.
- Allow for integrated use of heterogeneous resources from iVDGL, European DataGrid (EDG) [3] and CrossGrid [4] testbed domains.

Many other projects have collaborated actively to this initiative: EDG has provided the middleware for collective services such as the EDG Resource Broker (RB) and the Data Management (DM) services together with some hardware resources for the realization of the testbed; the CrossGrid project in Europe has contributed with four sites; the CMS and Atlas collaborations have made some of their hardware resources available for the realization of the transatlantic testbed. In Figure 1 we show the geographical distribution of the testbed: in total ten sites in Europe and nine in the U.S. were part of WorldGrid.

In what follows we describe the architecture of the WorldGrid testbed, the problems encountered and the solutions taken in realizing it. In particular, we describe the problems related with the authentication and authorization mechanism; the Grid resources description via the GLUE (Grid Laboratory for a Uniform Environment) schema [8]; the job submission mechanism that allows for user requirement match-making through a DataGrid GLUE-aware Resource Broker successfully extended to U.S. sites; replica management and for CMS applications through DataGrid middleware; and finally the creation of a coherent and automatically installable software distribution that can satisfy existing deployment tools such as LCFG (EDG Local Configuration System for farms) [9] used by DataGrid in Europe and PACMAN [11] used by the Virtual Data Toolkit (VDT) [10] in the U.S.. In order to monitor the behavior and performance of WorldGrid, the EDT-Monitor package has proven to be a good prototype to be deployed at operation centers. Through a geographical map of the sites, an operator has access to the resource and services status and can spot problems on-line.

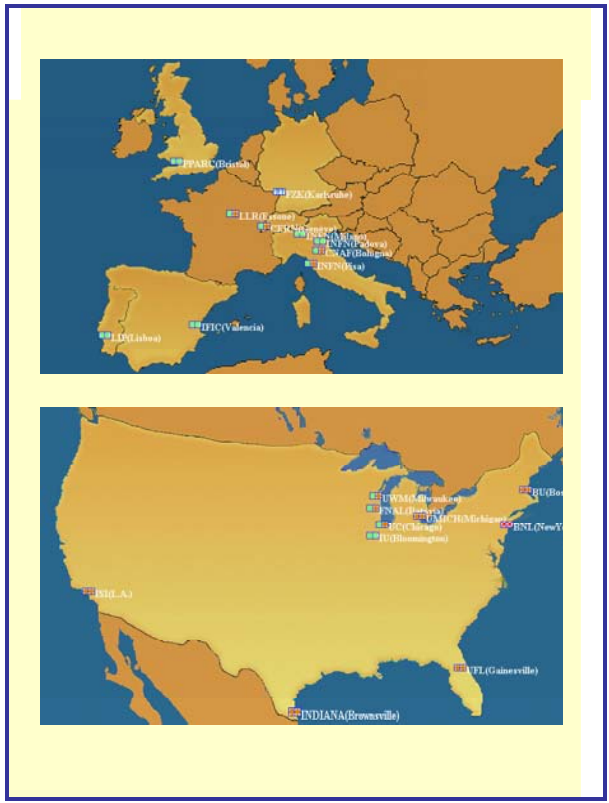

**Figure 1: The WorldGrid testbed: EU and U.S.**

## 2.  GRID MIDDLEWARE AND DOMAINS

The initiative began with a study of the basic services and architecture in the European and American Grid testbeds. Both Grids were built on top of basic Globus services, however the architecture was different. Here we list the main differences we had to deal with in order to realize interoperability between the Grids.

### 2.1.  The European DataGrid Testbed

The European DataGrid Testbed is a homogeneous Grid where the following software was installed and uniformly used at each site:
- RedHat 6.2 operating system (OS)
- Basic external packages needed
- Globus software (Globus 2.0 + EDG specific patches).

The following basic services of Globus are used: the Globus Security Infrastructure (GSI); the Information System based on the Globus Monitoring and Discovering Service (MDS), the Globus Job Submission Services based on the GRAM (Globus Resource Allocation Manager) service, the Globus Gatekeeper and the Globus Jobmanager; the Globus Replica Catalog based on LDAP and the Replica Management Services. On top of these components, DataGrid has built optional collective services to provide a more complete functionality.

In Europe DataGrid recognizes a list of well-established Certification Authorities (CAs) which sign host and user certificates. Such certificates are registered with a Virtual Organization (VO) LDAP server. Starting from these VO servers, the grid-mapfiles of the various computing and storage resources are populated, giving access to VO users.

Resources are organized for the services provided. Computing Elements (CEs) are farms of computing nodes managed by a head node where the Globus Gatekeeper and a Local Batch Resource Manager (LRM) run. In the resource description, a CE is mapped to an LRM queue.

On the Storage Elements (SEs) the Globus and EDG Data Management Service run to serve user requests for data replication and registration in the central Globus Replica Catalogue (RC).

The Information Service (IS) reports the status of all Grid resources collected from the local Grid Resource Information Services (GRISes). The information is organized in a schema defined appropriately by EDG. Such information is used by the EDG Resource Broker to match user requests against available resources.

The EDG Resource Broker (RB) is responsible for receiving user requests for job submission and finding the best suitable resource to execute the job according to user requirements.

The EDG Logging and Bookkeeping Service keeps track of job execution and status.

An EDG Proxy server takes care of user proxy renewal.

The last component is the User Interface (UI) that is the portal to the European DataGrid testbed. On the UI, the user has a command line interface or a Grid portal for job submission to the European DataGrid.

In Figure 2 we show the basic components of the EDG testbed, the flow of the matchmaking process, the job submission and replication requests.

As last note, the software distribution, installation and configuration of the EDG software is quite automatic, providing support for both centralized farm installation and for desktops. The LCFG EDG component allows system administrators to control the entire installation and configuration of a Grid site from a central point describing in a database and via LCFG configuration objects/modules the functionality of the nodes at a site.

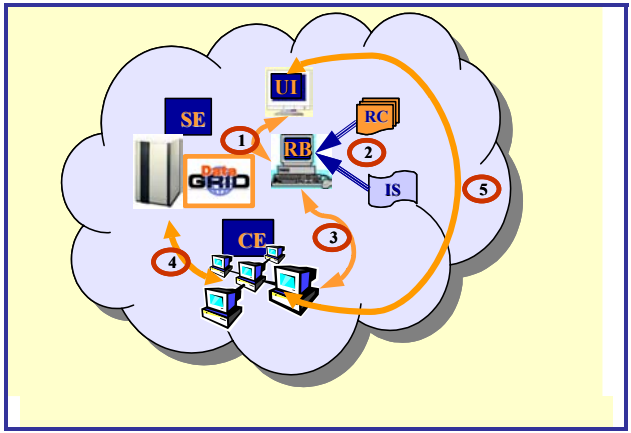

**Figure 2: Basic EDG components and job flow**

## 2.2. The iVDGL Testbed

Before the WorldGrid initiative, in the U.S. the architecture of the iVDGL testbed was quite different from the one in Europe, even if the basic Grid services used were the same ones from the Globus Toolkit. First of all, the testbed was very heterogeneous in terms of operating systems (RedHat 6.1, 6.2, 7.3, FermiLinux 6.1, etc.), external packages and even Globus versions (2.0 and 2.2) adopted. The Grid software distribution used was the Virtual Data Toolkit (VDT) that includes the Globus Toolkit, a distribution of the Condor and Condor-G [12] software and some of the data management tools developed in collaboration with EDG, such as the Grid Data Mirroring Package (GDMP) [13]. The VDT distribution is organized in two main components: a VDT Server and a VDT Client. The Server part of the distribution includes all Globus services available: GSI, MDS, Globus Schema for Resource Monitoring and Discovery, GRAM, Globus Gatekeeper, Globus Jobmanager, Globus and GDMP Replication Services, Local Resource Management Services (Condor). An iVDGL VDT Server could be viewed as an EDG CE and SE all together.

Through the Globus Schema, resources were described and published in the MDS for further monitoring.

A VDT Client corresponds more or less to an EDG UI. Through the VDT Client a user could access the iVDGL testbed, submit jobs, move data through the Globus GASS and GridFTP mechanisms.

A user authenticated him/herself on the iVDGL Grid via his/her PKI certificate signed by the Globus Certification Authority. The mapping between user certificates and local accounts on the iVDGL testbed was done by hand populating with an appropriate list of certificates the local grid-mapfiles on the machines providing the services.

In order to execute a job, it is the responsibility of the user or of the application infrastructure to find the best resource to run the job. A component such as the EDG RB is missing in the iVDGL testbed. Usually a round-robin mechanism or direct submission to a specific resource is applied in the iVDGL testbed. In addition, data movement between the VDT client and the execution machine has to be performed in advance by the user or the application.

In Figure 3 we show the architecture of the iVDGL testbed and the job flow.

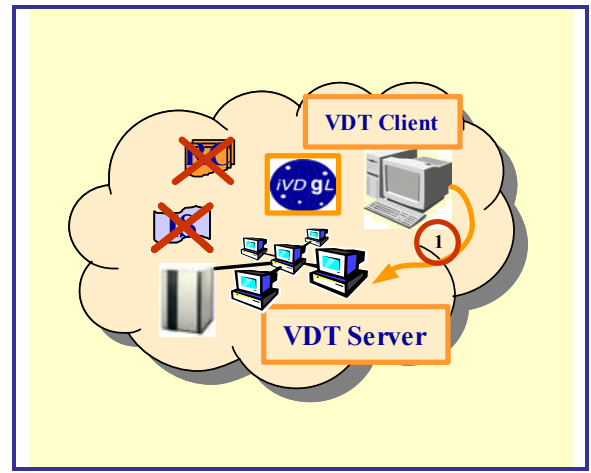

**Figure 3: Basic iVDGL components and job flow**

## 2.3. The WorldGrid Testbed

The WorldGrid testbed software distribution is based on the middleware provided by the EDG 1.2 release for European sites and the VDT 1.1.3 release for U.S. sites. The final configuration consisted of one central EDG-Resource Broker (RB) at INFN-Pisa with a backup at INFN-Milan; one central EDG-Information Index (II) at INFN-Pisa with a backup at INFN-Milan; one central EDG-RB GLUE-compliant at INFN-CNAF (Bologna); one central Globus Replica Catalogue Server (RC) at INFN-CNAF; two EDG-User Interfaces (UIs) running the GENIUS portal at INFN-Padova and INFN-Pisa and two UIs running the Grappa portal at ANL and INFN-Milano; one Virtual Organization (VO) LDAP server at INFN-CNAF for DataTAG authorized users, and one at ANL for iVDGL authorized users; 8 European sites hosting EDG Computing Elements (CEs), Worker Nodes (WNs) and Storage Elements (SEs): Bologna, Milano, Padova (INFN-Italy), Valencia (Spain), Geneva (Switzerland), Bristol (UK), Karlsruhe (Germany) and Lisbon (Portugal); 9 U.S. sites hosting VDT servers and clients at Gainesville (FL), Batavia (IL), Bloomington (IN), Boston (MA), Milwaukee (WI), San Diego (CA), Pasadena (CA), Argonne (IL) and Brookhaven (NY). As Local Resource Manager System (LRMS), PBS was adopted by 7 European sites and 5 U.S. sites, LSF by 1 European site and Condor by 4 U.S. sites.

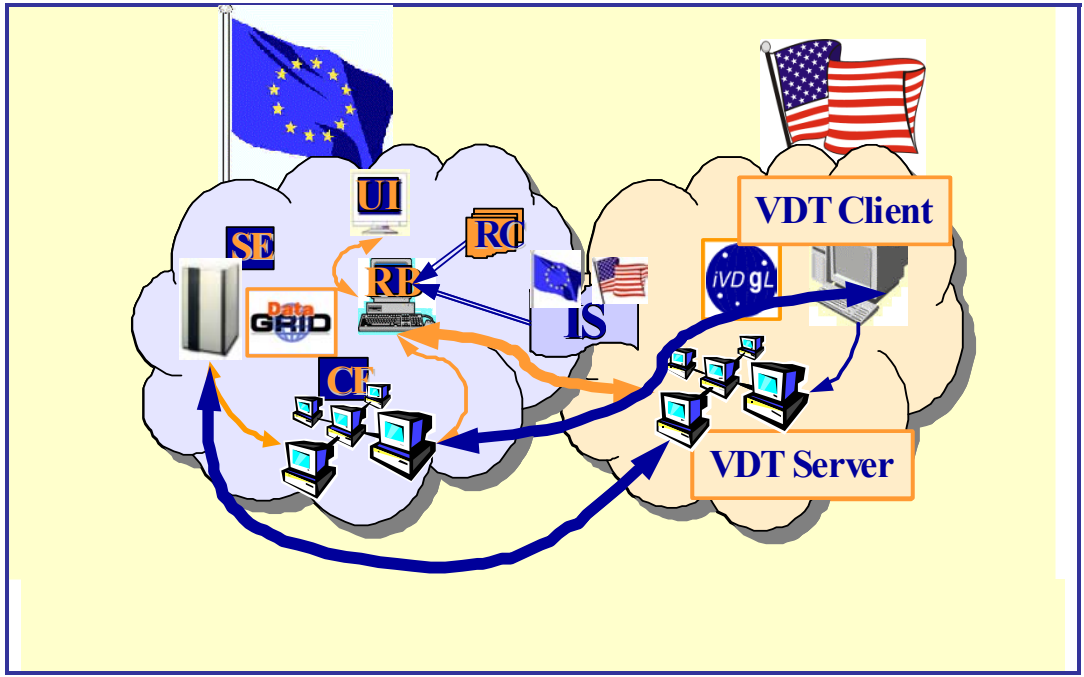

**Figure 4: Final architecture of the WorldGrid testbed**

The operating systems supported are RH6.2, Fermi 7.1 and RH7.2.

In order to achieve interoperability between the two testbeds, i.e. allow for job submission through the EDG-RB to both European and U.S. sites, additional software has been identified and installed layering it on top of the existing EDG and VDT installations:

- VO management tools for authentication/authorization: while these tools exist in the EDG 1.2 distribution, they are missing in VDT-1.1.3;
- Information Services: the Globus Schema and Information Providers (IPs), part of VDT-1.1.3, are not descriptive enough for the resource discovering and brokering process performed by the EDG-RB. EDG and Glue Schema and Information Providers were installed in addition;
- Data Management: some EDG-1.2 tools used by ATLAS and CMS applications for running their jobs and managing their input/output data on the EDG testbed were installed on VDT sites;
- Monitoring: two different monitoring tools, Nagios and VO Ganglia have been adopted, and the corresponding packages installed and configured on all sites.
- Applications: ATLAS and CMS jobs assume to find part of the experiment specific software installed on the WN where they are executed. The experiment software distribution is already included in the standard EDG-1.2 release, and had to be installed on VDT servers;

By the time of the IST2002 demo, a total of 8 European sites and 5 U.S. sites were properly set up and could be selected by the EDG-RB for job submission, fully exploiting its resource discovery and match-making capabilities. VDT servers were configured to act as CE and SE at the same time, while VDT clients acted as WN in the EDG language. ATLAS and CMS application jobs could run indifferently on European or U.S. side.

Moreover, INFN-Milano, INFN-Padova and Gainesville (FL) sites installed GLUE compliant schema and IP, and could be selected by the GLUE-compliant EDG-RB deployed in time for SC2002 demo.

**THCT006**

## 3. AUTHORIZATION / AUTHENTICATION TOOLS

In order to be able to transparently submit jobs between the two Grid infrastructures in the U.S. and Europe, one of the first problems encountered was authorizing and authenticating users in a uniform way.

On the iVDGL testbed, the Globus Security Infrastructure (GSI) is adopted: the Globus Certification Authority used to sign X.509 host, service and user certificates that were accepted throughout the testbed. On the EDG testbed, the Globus CA was not trusted since it did not comply with the EDG rules for a Certification Authority. In fact, very little was done to really ensure the identity of a user before signing his/her certificate. In order to overcome this problem, the U.S. Department of Energy (DOE) has established an EDG compliant Certification Authority that was therefore trusted by EDG. Together with DOE signed certificates, on the iVDGL testbed the certificates signed by most of the EDG trusted CAs had to be accepted in order to allow for transparent use of the iVDGL testbed by the EDG user community. Therefore, in the WorldGrid software distribution, the proper CA files (including the CA certificates, certificate revocation lists parameters, etc.) were included for both DOE and EDG trusted CAs.

For what concerns authorization, in the iVDGL testbed local Globus grid-mapfiles were populated manually with an appropriate well-established list of user certificates mapped to local accounts. In EDG, instead, support for a Virtual Organization (VO) environment is given. A user is registered with a VO and his/her certificate appears listed in a central VO LDAP server. Starting from these servers, through the mkgridmap mechanism, the local grid-mapfiles installed on the machines throughout the testbed are automatically populated with the certificates of the users listed in the VOs "supported" by the machines at a given site.

In the WorldGrid testbed, we created two VOs: the iVDGL and DataTAG VOs. Users registered with one of these VOs could use the WorldGrid testbed. The EDG mkgridmap mechanism has also been adopted in the iVDGL part of the WorldGrid testbed and mkgridmap made part of the WorldGrid software distribution.

Many sites participating to the WorldGrid initiative had their own authorization mechanism in place that was sometimes not compatible with GSI. This is the case for FermiLab, where a very secure infrastructure based on Kerberos V is in place. Because of a missing interface between GSI and the various existent infrastructures, we had to skip the local mechanism and only rely on GSI.

Certification Revocation Lists (CRLs) were kept update by running appropriate scripts in cron jobs on all Grid resources. In order to have this mechanism work, we had to ensure that the update scripts were installed and executed in a cron on all Grid resources.

The last problem we encountered concerned opening the use of site and laboratory resources to WorldGrid registered users. Normally, sites have internal policies to guarantee access to their resources. Sometimes users have to go through special training and sign specific policy agreements in order to use the computing resources at a site. In EDG, all sites part of the collaboration agree to open their resources to the users who sign the EDG usage policy statement. Something similar had to be done for WorldGrid.

## 4. RESOURCE DISCOVERY AND SHARING

If different testbeds in different Grid domains have to co-operate, they need to be aware of each other and be able to share their resources. In order to achieve this goal and set up the WorldGrid testbed we had to take some actions on the heterogeneous machines of the U.S. and European sites. A central information system has been put in place: both U.S. and European resources registered with the top Globus GIIS (Grid Information Index Service) in Pisa, Italy.

### 4.1. From U.S. to Europe

For the U.S. testbed to harvest European resources, small effort was needed. Actually, we had to install no extra middleware components beside the existing standard EDG software. This is because the resource discovery is not yet fully automated in the U.S. testbed, and the target machine for a job is specified by hand at job submission time. Having a Globus gatekeeper running on a machine – which EDG has since it is based on Globus like VDT – and granting GSI access rights was enough to assure sharing of that resource.

### 4.2. From Europe to the U.S.

Vice versa, in order to choose a suitable machine for a job, the European testbed relies on the Resource Broker, which in turn gets the needed information from the Information Service.

The Information Service is a tool designed to support the initial discovery and ongoing monitoring of the existence and characteristics of resources, services, computations, and other entities, and is a vital part of a Grid system [15]. The EDG testbed adopted the Globus implementation, called MDS-2, which is based on the LDAP model and uses OpenLDAP in addition to some custom database back-ends[17].

The LDAP data model represents information as a set of objects organized in a hierarchical namespace. Each object is tagged with one or more named types and contains bindings of values to named attributes according to the object type. Definitions of object types are called *schema*. Actual values for the attributes are provided both statically by configuration files and dynamically by software components (usually Perl scripts) called *information providers* [16,18].

**THCT006**

When first released, the Globus MDS schema was proven by EDG to provide insufficient information for an efficient resource brokering. Then new schemas were defined and the relative information providers were developed in EDG [25]. Being not in the standard Globus distribution, these components were missing on the machines of the U.S. testbed. Since they are essential in order for the RB to "see" a resource and take it into account in its brokering process, we had to find a way to install them on U.S. machines.

Our first task was to isolate them from their EDG environment. While the information providers themselves are not tightly tied to other EDG components, their configuration process relies highly on operations performed by LCFG, the EDG installation tool. LCFG takes the required information from a lot of "profile" files describing the complete configuration of a machine, so we had to modify the configuration files structure to be more self-contained.

Moreover, it was important to have in WorldGrid installation and configuration methods for the U.S. sites that minimize the need of human intervention. In order to satisfy this requirement, we gathered in a text file all the information that a site administrator has to provide, in the form of key/value pairs; then, by means of a set of scripts (largely based on EDG code) we make use of this information to create and fill up the required configuration files. One major issue was to integrate these files in the local MDS system preserving the existing configuration, which could be, in principle, used by other non- WorldGrid related programs. Since on many machines there was more than one single configuration, human intervention was still needed to identify the correct one: for this task, the best we could do was to provide detailed instructions for the site administrator.

After some difficulties with the very first machines, we refined both the scripts and the instructions in a cycle of effective co-operations with the most active members of the iVDGL project. At the end, all the participating sites were able to install and configure the information providers with small effort, and all the machines became visible to the Resource Broker.

At a second stage, we integrated as well the common GLUE schema and information providers defined and developed inside the iVDGL and DataTAG collaboration for a common description of the resources. Such components were installed on both the European and the U.S. resources and a GLUE-aware version of the EDG Resource Broker was also deployed to make use of the new description.

### 4.3. About MDS-2

Trying to let the EDG information providers and schema coexist with the Globus ones, we discovered two minor issues in the current MDS implementation. The first one concerns the *ldif* files, i.e. the files used to feed the database backend with data. These are simple text files read by the database engine to create instances of the objects defined by the schemas, and they have to be specified in the LDAP configuration file. Apparently, more than one ldif file can be specified (we had one for Globus objects and one for EDG objects), but actually only the last one is really used, due to the design of the parsing algorithm for the configuration file. This caused some information to disappear from the LDAP database, and the only work-around was to merge the two ldif files to a big one and leave it as the only one in the configuration file. We disliked this solution, because it was not clear who was in charge of managing this file, and because synchronization problems could arise if both EDG and Globus were to make changes to it.

A second issue revealed when we tried to run some queries against particular distinguished names: the LDAP engine incorrectly returned also objects whose distinguished names ended with the requested one (e.g., "hostname=grid001" matched also for "mds-hostname=grid001"). Fortunately, this did not affect the queries of the Resource Broker, which has a filter on the object class for the information returned by the Information Service.

## 5. JOB MANAGEMENT AND SCHEDULING

In the WorldGrid testbed the Workload Management System (WMS) software from the EDG project was adopted as a collective service.

The WMS components are the following: User Interface (UI), Resource Broker (RB), Logging and Bookkeeping (LB) Service, and Job Submission Service (JSS).

The UI [22], providing a language to describe characteristics, requirements, preferences of a job and a set of commands to manage jobs on the Grid, allows a user to interact with the Grid in order to perform operations such as submitting jobs, controlling their execution, and retrieving their output.

The RB [21] is the core component of the WMS, its main task is to find a Computing Element (CE) that best matches the requirements and preferences of a submitted job, considering also the current distribution of load on the Grid. Additionally, the RB allows for canceling a job and retrieving the output once a job has completed.

After consulting the Information Service and the Replica Catalogue for resolving data location, a suitable resource is found. The job is then passed to the JSS [20] for the actual submission to the selected CE. One of the JSS tasks is also to monitor the job execution and react if an event affects the job.

The LB Service [19] collects information about the scheduling system (*logging information*) and about active jobs (*bookkeeping information*) from the other components of the WMS.

**THCT006**

## 5.1. Job Submission through the EDG Resource Broker

In the WorldGrid testbed, two central RBs were deployed: one is an installation using the standard EDG software environment on a PC in Pisa, Italy, and the other is a VDT client installation with the RB, JSS and LB components from EDG on a PC in Milan, Italy. The second installation was done to prove that it was possible to integrate the EDG WMS components in the VDT environment. No major problems were encountered if the RedHat 6.2 release of the OS was used.

The Resource Information was mirrored on the Information Servers in Pisa and in Milan. Thus, each instance of the EDG RB pointed to the local Information Server making the two systems independent and redundant. All WorldGrid resources were available for job submissions from the two RBs.

Later on, another EDG RB was modified to make it GLUE-aware: certain GLUE attributes for ranking were recognized by this new broker. Few resources in WorldGrid in U.S. and Europe were publishing their status also through the GLUE schema in the Information Service. In the way, the GLUE-aware RB could execute matchmaking on them.

Together with the two RBs, three UIs were available: one in Pisa, one in Milan and the last one in Padova.

On the WorldGrid testbed job submissions could be performed in several ways: from one of the UIs, through the RBs, jobs could be sent for execution either by specifying a particular resource or by the broker matchmaking mechanism. U.S. jobs were sent directly to a EU or a U.S. resource for execution via direct Globus submission commands. Using the RB, a job could first stage its executables and files on the execution machine (as it was the case for U.S. ATLAS applications) and then run, or select a machine with a preinstalled environment and run there the specific application.

## 6. DATA MANAGEMENT

On the iVDGL testbed, the data management middleware tools available were mainly the ones coming from Globus. It is up to the application to take care of pre-staging application files or move/replicate data files in place before executing a job. This is normally done using the globus-url-copy command or the GridFTP client/server interface. The Globus Replica Catalogue is not used by the U.S. applications that store file locations in their private databases. The CMS experiment has used the GDMP tool for automatic file replication between sites, not using however the feature of automatic registration of files in the Globus Replica Catalogue.

EDG has developed quite few tools to make data handling easier on the Grid: the edg-replica-manager allows users to replicate files point-to-point between a UI, a CE or a WN and an SE. The edg-rc-server allows for setting up easily a Globus Replica Catalogue in a VO.

All these tools were additionally installed on the iVDGL sites. Sometimes this implied a porting to a slightly different version of the OS/compiler, but no major problems were encountered. Through these tools, EU applications could easily move the data produced in central servers and replicate them on other SE for further processing.

Outbound connectivity on the worker nodes or computing resources was required to copy Input/Output Sandboxes from/to EDG Worker Nodes or U.S. computing resources to/from the EDG Resource Broker. It is also required for update to the Replica Catalogues as well as for third party transfers.

For some CMS applications it was also necessary to run globus-job-run commands to create directories in the job execution script, which requires also inbound connectivity on given TCP ports.

## 7. PACKAGING

One of the big issues still open today was packaging and distribution of the software. The EDG machines in Europe part of the WorldGrid testbed were installed automatically using the LCFG tool. Such a tool is built around the RedHat RPM system. All packages to be installed on a Grid farm have to be distributed in RPM format and configured via LCFG configuration modules. Local configuration parameters are stored in a central database at a site, used to bootstrap the installation system. The installation and configuration steps are separated to allow for reconfiguration of single package components. Because EDG foresees the installation of packages even at the level of a WN, the LCFG tool turned out to be quite powerful for farm management.

In U.S. the PACMAN system is used to install and configure VDT Servers and Clients on several platforms. Packages can be distributed in any packaging format: tar files, cpio, RPM, etc. A PACMAN description file describes the software source, installation and configuration steps that can be system specific and executed in a single atomic operation. A PACMAN cache contains the software distribution consisting of the PACMAN description files associated with it.

In order to create a WorldGrid software distribution, we had to package and distribute the software using RPMs and LCFG objects and creating an appropriate PACMAN cache. Some tools to automate the process of creating RPMs starting from the PACMAN description files and vice versa where studied. Also the configuration steps between the two distributions had to be made consistent. Wherever possible a PACMAN description file used, as source, the correspondent package RPM. However, this imposed the package to be installed as root, something that was not always accepted by the site administrators.

From the WorldGrid effort another GLUE initiative has started to solve the installation and configuration issue in a common and coordinated way. For more details, see [27].

**THCT006**

Another GLUE initiative started with the WorldGrid experience is the one concerning the validation and certification of a site that wants to join an existing Grid. Inside WorldGrid we setup a set of tools that a site could run in order to verify that the installation was correct. Further functionality check of the entire Grid was executed via the monitoring tools.

## 8. MONITORING TOOLS

Once the WorldGrid testbed was up and running and executing application jobs, three monitoring tools developed especially for WorldGrid were used: one is MapCenter developed by EDG, one developed in Europe by DataTAG based on the Nagios technology and the third one developed in U.S. based on Ganglia.

MapCenter has been tailored to monitor the status of specific services for the WorldGrid testbed.

The Ganglia tool allows for Grid and VO level views of resource activities, utilization and performance. Data is collected from other host-level monitoring tools (e.g., gmond, rocks, nagios-exporter) or even other ganglia daemons. Data is manipulated in XML format and stored locally where the central demon runs in a RRD for statistics purposes. Host-level collectors (gmonds) are used on remote nodes to collect information for load, memory, swap, disk, network and user-defined metrics.

The tool based on Nagios permit nice views of the status of the testbed that are selectable by VOs, countries and sites. Various Nagios plug-ins were developed. Sites and corresponding metrics can be represented over a world map. Dots with different colors are displayed to display the metric of a site.

Details about the Nagios tool developed by DataTAG can be found in this conference's proceedings [26].

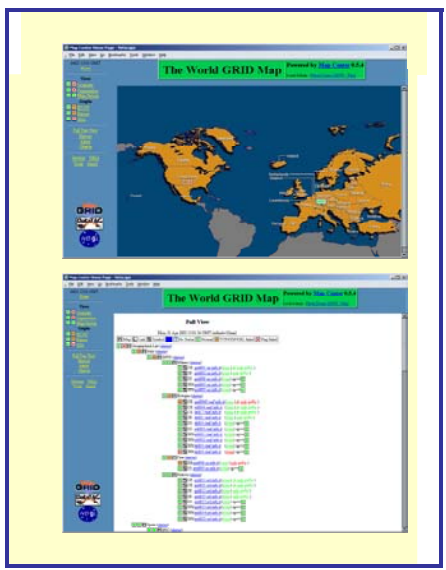

**Figure 5: The WorldGrid MapCenter**

## 9. HEP APPLICATION EXECUTION AND WEB PORTALS

Two LHC experiments, ATLAS and CMS, successful exploited the WorldGrid testbed for executing jobs simulating the response of their detectors to physics events expected from the real collisions that will happen at LHC starting from 2007.

Jobs of this kind are needed in order to optimize the detector design and performance, and to study the algorithms that better filter the rare interesting physics events the experiments aim to detect. They produce a large amount of data, and since many years they run on local dedicated computing farms consisting of hundreds of nodes and Terabytes of disk and/or tape storage. Within the WorldGrid testbed, for the first time HEP simulation jobs were submitted and run indifferently on U.S. and European farms, producing data which could be retrieved and further analyzed on the submitting machine, or simply stored on the remote farm where they were created and registered on a Replica Catalogue on the Grid which made them available for further processing.

Job submission was performed through two different web portals, GRAPPA for ATLAS and GENIUS for both CMS and ATLAS. The advantages of using web portals for job submission over the Grid are well known: they can nowadays be accessed from everywhere and from various devices (desktop PC's, laptops, PDA's, cell phones…); they present the same user interface to several back-ends, preventing the user from learning different Grid "dialects"; they have user-friendly layout and make many Grid services just one mouse click away.

### 9.1. GRAPPA/ATLAS job submission

The GRAPPA portal (**GR**id **A**ccess **P**ortal for **P**hysics **A**pplications) has been developed within the Grid Physics Network Project (GriPhyN [14]), the U.S. ATLAS Software and Computing Project, and the International ATLAS Experiment, with the goal of providing ATLAS physicists a point of access to Grid resources. It was installed in a U.S.-VDT server of the WorldGrid testbed, and allowed ATLSIM job submission to all U.S.-VDT servers and EU-EDG sites included in the testbed. The portal uses the Grid Information System and gives access to the list of available resources, resource usage statistics, monitoring tools and accounting information for all resources on the Grid. The user fills in a web form job parameters specifications and select the remote resource where to submit the job. He can check the status of the job as it is running, and at the end retrieve the output data. EU-EDG remote resources are reached submitting the job to an EDG-UI Grid element, which in turn submits the job to an EDG-RB which selects the best suited EDG-CE for running the job, performing a match between the job requirements and the CE characteristics and workload information.

### 9.2. GENIUS/CMS-ATLAS job submission

The GENIUS portal (**G**rid **E**nabled web e**N**vironment for site **I**ndependent **U**ser job **S**ubmission) has been developed by INFN and NICE SRL in Italy, within the INFNGRID project. It is installed on a EDG-UI and allows users belonging to any Virtual Organization for:
- Secure authentication and authorized access to the Grid;
- Job submission and job status monitoring on the Grid;
- Monitoring of Grid resources;
- Data management over the Grid (upload, download, replication and publication of files)
- Replica Catalogue browsing
- Interactive sessions on remote EDG-UI

The JDL [23,24] file describing the user job can be uploaded to, or directly edited on the remote EDG-UI hosting the portal. In the case of ATLSIM jobs, some additional input files are uploaded to the EDG-UI, and specified in the JDL file. GENIUS allows the user to select an RB among a list of available ones. The EDG-RB, after having received the job, performs a match-making between the job requirements and the resources characteristics, allowing the ATLSIM job to be scheduled only in the farms publishing the ATLAS tag (it means they have certified ATLAS software installed) and where the Grid user is authorized to run his jobs. In the WorldGrid testbed, all demo users were authorized to run on all resources at the participating sites and all sites were installed with certified ATLAS software, so the job could run everywhere in U.S. or EU sites. At the end the output could be retrieved on the submitting EDG-UI, and immediately analysed through the portal via its VNC interface allowing the user to open an interactive graphic session on the EDG-UI.

This approach could be adopted for every VO, once the VO-specific job description is translated into a JDL file. Nevertheless, the GENIUS portal allows for a more VO-specific customisation, defining the so-called "VO Services" item.

This is the case of CMS, which had since time a set of MonteCarlo production tools (IMPALA, BOSS, RefDB) developed for automatic simulations jobs submission and tracking on the local farms. These tools were interfaced to GENIUS portal, allowing for a user-friendly configuration from any remote web browser. All EDG-UIs used in the WorldGrid testbed were equipped with these additional CMS software components.

CMS applications allowed also to fully exploit in the WorldGrid environment the EDG data management tools like the EDG-RC, and the capability of the EDG-RB to select the Grid resources, querying the EDG-RC, according to physical replica locations of input data required by the job for processing.

A more detailed description is given in the paper in these proceedings [28].

## 10. CONCLUSIONS

The WorldGrid testbed allowed exploiting the possibility of interoperable co-operating Grids offering transparent access to users. Such a Grid can only be possible when the basic middleware realizing the key services is the same on all Grids. WorldGrid offered as well the possibility of understanding what are the open issues to address in order to reach basic interoperability: authentication and authorization, resources description and status, site testing, validation and monitoring, packaging and distribution. Another important aspect is co-operation and policy agreement among the participating sites.

The experience acquired during the WorldGrid experience and even some of the developments are currently used by LCG for the realization of a Grid worldwide service.

### Acknowledgments

The authors wish to thank the following people for their valuable contribution to this work:
**DataTAG:** S. Andreozzi, R. Barbera, S. Fantinel,
      A. Ghiselli, M. Mazzucato, G. Tortone,
      C. Vistoli – INFN, Italy
**DataGrid:** M. Draoli - CNR-Rome, Italy
      E. Ronchieri - INFN, Italy
      H. Stockinger - CERN, Switzerland
**CrossGrid:** A. Garcia, M. Hardt, FZK - Germany
      J. Marco, UC - Spain
      M. David, J. Gomes, LIP - Portugal
      O. Maroney, U.Bristol, UK
**iVDGL:** P. Avery, J. Rodriguez - U. of Florida
    E. Deelman, N. Olomu - USC/ISI
    J. Gieraltowski, S. Gose, E. May, J. Schopf – Argonne
    A. Afaq, J. Annis, R. Glossum, R. Pordes, V. Sekrhi – FermiLab
    W. Deng, J. Smith, D. Yu - BNL
    A. DeSmit, A. Roy - Wisconsin
    C. Dumitrescu, I. Foster, R. Gardner - U. Chicago
    L. Grundhoefer, J. Hicks, F. Luehring, L. Meehan - U. Indiana
    S. Youssef, Boston University
    B. Moe - Milwaukee
    D. Olson – LBNL
    S. Singh - Caltech

Work supported by the European Commission EU Grant IST-2001-32459.

*CHEP 2003, La Jolla – CA, USA, March 24-28 2003* 10## References

[1] European DataTAG Project: http://www.datatag.org
[2] iVDGL Project: http://www.ivdgl.org
[3] EU DataGrid: http://www.eu-datagrid.org
[4] CrossGrid: http://www.crossgrid.org
[5] Globus Project: http://www.globus.org
[6] IST2002: http://europa.eu.int/information_society/programmes/research/ist_event_2002/index_en.htm
[7] SC2002: http://www.sc2002.org/
[8] GLUE Schema: http://www.cnaf.infn.it/~sergio/datatag/glue/index.htm
[9] EDG Local Configuration System (LCFG): http://hep-proj-grid-fabric.web.cern.ch/
[10] Virtual Data Toolkit (VDT): http://www.lsc-group.phys.uwm.edu/vdt/
[11] PACMAN: http://physics.bu.edu/~youssef/pacman/
[12] Home Page for the Condor Project: http://www.cs.wisc.edu/
[13] GDMP: http://www.cern.ch/GDMP
[14] GriPhyN: http://www.griphyn.org
[15] Grid Information Services for Distributed Resource Sharing, K. Czajkowski, S. Fitzgerald, I. Foster, C. Kesselman. *Proceedings of the Tenth IEEE International Symposium on High-Performance Distributed Computing (HPDC-10)*, IEEE Press, August 2001.
[16] A directory service for configuring high-performance distributed computations, S. Fitzgerald, I. Foster, C. Kesselman, G. von Laszewski, W. Smith, and S. Tuecke. In *Proc.6th IEEE Symp. on High Performance Distributed Computing*, pages 365–375, 1997.
[17] Usage of LDAP in Globus, I. Foster, G. von Laszewski. Draft, 1998 - ftp://ftp.globus.org/pub/globus/papers/ldap_in_globus.pdf
[18] Time, Information Services and the Grid. Brian Coghlan, Abdeslem Djaoui, Steve Fisher, James Magowan and Manfred Oevers. *DataGRID EDMS Publications,* 2001 - https://edms.cern.ch/document/349802/3
[19] The Logging and Bookkeeping Architecture http://www.cnaf.infn.it/giaco/wp1/lb_draft.pdf
[20] Job Submission Service Architecture and APIs http://www.pd.infn.it/~sgaravat/Grid/jss-arch.pdf
[21] Resource Broker Architecture and APIs http://www.infn.it/workload-grid/docs/20010613-RBArch-2.pdf
[22] Job Submission User Interface Architecture http://www.infn.it/workload-grid/docs/DataGrid-01-TEN-0103-0_0-Document.pdf
[23] Job Description Language How-To http://www.infn.it/workload-grid/docs/DataGrid-01-TEN-0102-0_0-Document.pdf
[24] Job Submission User Interface man Pages http://www.infn.it/workload-grid/docs/DataGrid-01-TEN-0101-0_1-Document.pdf
[25] WP1 Inputs to the DataGrid Grid Information Service Schema Specification http://www.infn.it/workload-grid/docs/wp1-gis-gos.pdf
[26] A monitoring tool for a GRID operation center, S. Andreozzi, S. Fantinel, D. Rebatto, L. Vaccarossa, G. Tortone, CHEP2003 Proceedings PSN MOET002
[27] A GLUE Meta-Packaging proposal for Grid Middleware and Applications: a joint project between LCG, EDG and iVDGL, O. Barring, G. Cancio, F. Donno, S. Youssef, J. Rodriguez, A. Roy https://edms.cern.ch/file/385786/1.2/GLUE_Packaging.pd
[28] ATLAS and CMS applications on the WorldGrid testbed, V. Ciaschini, F.Donno, A. Fanfani, F. Fanzago, V.Garbellotto, M. Verlato, L. Vaccarossa, CHEP2003 Proceedings PSN TUCP004.**THCT006**